\documentclass[11pt]{article}
\usepackage{moriond,epsfig}

\def\Journal#1#2#3#4{{#1} {\bf #2}, #3 (#4)}

\def\NPB{{\em Nucl. Phys.} B}

\def\PRL{{\em Phys. Rev. Lett.}}
\def\PRC{{\em Phys. Rev. C}}

\def\nat{{\em Nature}}
\def\jetp{{\em J. Exp. \& Th. Phys.}}
\def\aap{{\em Astron. \& Astroph.}}
\def\aaps{{\em Astron. \& Astroph. Suppl.}}
\def\araa{{\em Ann. Rev.  Astron. \& Astroph.}}
\def\apj{{\em Astroph. J.}}
\def\adspr{{\em Adv. Space Research}}
\def\pasp{{\em Publ. Astron. Soc. Pacific}}
\def\mnras{{\em Mon. Not. R. A. S.}}

\def\be{\begin{equation}}
\def\ee{\end{equation}}
\def\bea{\begin{eqnarray}}
\def\eea{\end{eqnarray}}

\def\simlt {\lower.5ex\hbox{$\; \buildrel < \over \sim \;$}}
\def\simgt{\lower.5ex\hbox{$\; \buildrel > \over \sim \;$}}

\newcommand{\msun}{M$_{\odot}$}
\newcommand{\tiff}{$^{44}$Ti}
\newcommand{\caff}{$^{44}$Ca}
\newcommand{\scff}{$^{44}$Sc}
\newcommand{\nifs}{$^{56}$Ni}
\newcommand{\cofs}{$^{56}$Co}
\newcommand{\fluxunit}{${\rm ph^{-1}cm^{-2}s^{-1}}$}
\newcommand{\sax}{{\it BeppoSAX}}
\newcommand{\osse}{{\it CGRO-OSSE}}
\newcommand{\integral}{{\it INTEGRAL}}
\newcommand{\comptel}{{\it CGRO-COMPTEL}}

\begin{document}
\vspace*{4cm}

\title{THE DETECTION OF THE $^{44}${\bf Sc} NUCLEAR DE-EXCITATION 
LINES AND HARD X-RAY EMISSION FROM CAS~A}

\author{JACCO VINK\footnote{Chandra Fellow}}

\address{Columbia Astrophysics Laboratory,  MC 5247, 
550 W 120th street, New York, NY 10027, USA}

\author{J. MARTIN LAMING}
\address{Naval Research Laboratory, Code 7674L, Washington DC 20375, USA}

\maketitle
\abstracts{
We discuss the detection of \scff\ (product of \tiff) 
in the supernova remnant Cas A with the \sax-PDS instrument, 
and elaborate on
the nature of the hard X-ray continuum, which is not only 
important for correctly estimating the \scff\ flux, but is interesting
in its own right. We apply the lower hybrid wave electron 
acceleration model to the hard X-ray data and use upper limits on
the cosmic ray injection spectrum to infer that $B > 2$~mG.
We conclude with the prospects of observing nuclear decay lines
and electron-positron annihilation in Cas A with \integral.
}

\section{Introduction}
Cas A is the brightest radio source, 
and one of the best studied supernova remnants.
With its likely explosion date of AD 1680, it is also
the youngest known galactic remnant.\cite{Ashworth80}
Optical and X-ray spectroscopy strongly suggest that its progenitor
was a massive star, probably a Wolf-Rayet star, which had
lost most of its mass by the time it exploded.\cite{Fesen87,Vink96}
The explosion was not recorded as a bright event, suggesting an
underluminous supernova.
Its radio brightness identifies Cas A as a source of cosmic rays,
and its youth makes it one of the best remnants to study the
explosive nucleosynthesis of massive stars.

In this paper we will relate to both topics, as the deep observation
of Cas A with \sax\ discussed here resulted in the
first detection of \scff\ nuclear de-excitation emission, 
confirming the synthesis of a substantial amount \tiff\ by the explosion.\cite{Vink01}
In addition the deep observation constrains
the properties of the hard X-ray continuum emission.

\begin{figure}
   \hbox{
\parbox{95mm}{\psfig{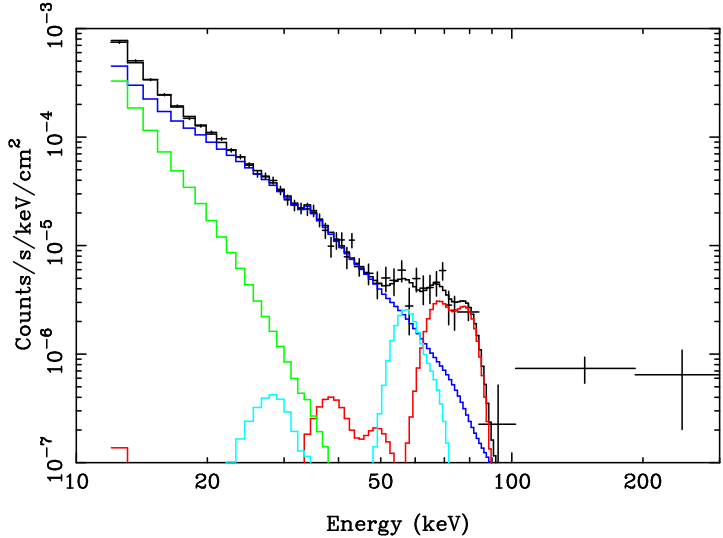}
	\vfill
	}
		\hskip 5mm
\parbox{50mm}{
\caption{\sax-PDS spectrum of Cas A with the best fit \scff/non-thermal 
bremsstrahlung model. 
The individual emission components are:
\scff\ line emission (red), 
non-thermal continuum (blue), thermal continuum
(green) and possible line contamination from collimator
material (Tantalum, cyan). The observed count rate in each channel 
has been divided by the effective area in order to yield approximate flux 
densities.
}}}
\label{fig:PDSspec}
\end{figure}

%\newpage

\section{The detection of \scff}
The composition of the inner most layers of supernova ejecta
is to a large extent determined by nuclear statistical equilibrium, 
\nifs\ being the most abundant element. However, as the layer expands
the density decreases and the triple-$\alpha$ reaction stops,
resulting in an excess of $\alpha$-particles;
a process referred to as alpha-rich freeze out.\cite{The98}
This condition favors the production of \tiff.
Although \tiff\ is much less abundant than radio-active \nifs, 
the longer decay time of 
$85.4\pm0.9$~yr\ \cite{Ahmad98,Goerres98,Norman98} makes
\tiff\ the main energy source for the expanding ejecta of some supernovae
$\simgt 2000$~days after the explosion, 
as is the case for SN 1987A.\cite{Diehl98,Lundquist02}
The observation of \tiff\ in young supernova remnants provides information
about the explosion that formed them,
as the amount synthesized depends sensitively on explosion energy, explosion asymmetries,
pre-supernova mass loss\footnote{A strong mass loss reduces late time fall 
back on the stellar remnant.}, and the
mass cut.\footnote{The mass cut defines the mass of the stellar remnant.
A massive stellar remnant will have accreted most of the \tiff\ (and \nifs)
produced.}

The decay chain is \tiff $\rightarrow$ \scff\ $\rightarrow$ \caff, and is
in more than 99\% of the decays accompanied by the emission of
three photons at 68 keV, 78 keV (from excited \scff) and 1157 keV (from
excited \caff).
Although Cas A is the best candidate for observing \scff\ and \caff\ emission,
the detection of \caff\ emission with \comptel\cite{Iyudin94} was 
a surprise, as the implied synthesized \tiff\ mass was large compared to
model predictions. Moreover, hard X-ray observations failed
to detect \scff\ line emission.\cite{The96,Rothschild97,Vink00}

To address this issue Cas A was observed in May and June 2001 for 500~ks
with the \sax, its hard X-ray experiment, PDS\ \cite{Frontera97}, 
being best suited
for detecting  \scff\ line emission, in part due to the low particle background
of the \sax\ orbit. 
With an additional 100~ks of archival data the resulting spectrum is
of high quality with an energy resolution 9~keV (FWHM) 
around 75~keV.

Fitting the data with a simple power law continuum model, shows
a significant excess at the energies of the \scff\ emission.
The excess can be fitted with gaussian lines at the \scff\ line energies,
with a flux for each line of $(2.0 \pm 0.4) \times 10^{-4}$~\fluxunit\ and a
combined detection significance of more than 5$\sigma$.\cite{Vink01}
For a distance of 3.4~kpc and an age of 320~yr this implies an initial
\tiff\ mass of $10^{-4}$\msun.
The measured power law slope is $\Gamma = 3.30 \pm 0.04$.
The \scff\ line flux is lower, 
but consistent with the latest \comptel\ measurements of
the \caff\ line flux of $(3.4 \pm 0.4) \times 10^{-4}$~\fluxunit.~\cite{Schoenfelder00}
However, the measured \scff\  line flux depends on  the
assumed shape of the continuum, whose nature is not yet established.

\section{The nature of the hard X-ray emission}
The hard X-ray emission of Cas A is interesting in its own right, 
and its existence is connected with the physics of electron heating and acceleration.
The hard X-ray component is likely to extend to lower photon energies, 
where it interferes with the accurate measurement of the 
thermal bremsstrahlung component, which
is important for inferring elemental abundances and temperatures.

The hard X-ray emission may be the result of
synchrotron emission from
extremely energetic cosmic ray electrons ($E \simgt 10$~TeV), similar
to what is now established to be the dominant X-ray continuum emission for 
SN 1006.\cite{Koyama95,Reynolds99} 
Alternatively, it can be the result of
bremsstrahlung from subrelativistic or mildly relativistic electrons,
in which case the electrons may constitute the cosmic ray injection 
spectrum, but not necessarily so.\cite{Asvarov90,Bykov99}

Here we highlight the application of a 
particular bremsstrahlung model,
which involves electrons accelerated by lower hybrid waves (LHW),
which have frequencies intermediate to the electron and ion  
gyrofrequencies.\cite{Laming01a,Laming01b}
LHW are excited in plasma in which the magnetized electrons are
in the presence of free streaming ions reflected by primary or 
secondary shocks.
They can explain the observation of energetic electrons at the earth bow 
shocks,
and X-ray emission from comets, and have in fact been observed 
in situ for Halley's comet.\cite{Bingham00}
LHW are promising for explaining the hard X-ray emission
from Cas A, as the magnetic field is relatively high 
($\sim$1~mG, see next section). The fact that 
Cas A is oxygen-rich is important, because it means that for a given shock velocity
the average ion is more energetic than for solar abundances.

The emission model consists of a thermal bremsstrahlung 
component of $kT_{\rm e} = 3.5$ keV and a bremsstrahlung component from accelerated 
electrons. The maximum electron energy obtained by the process, $E_{max}$, is
a free parameter.
The model fits the \sax-PDS data remarkable well up to 100~keV (Fig.~\ref{fig:PDSspec}). 
The data indicate  $E_{max} \sim 95$~keV and the emission measure of the
non-thermal component is 1/11th of that of the thermal component.
The model predicts a steepening of the spectrum near $E_{max}$.
For this model the measured \scff\ line flux is 
$(3.2\pm0.3)\times10^{-5}$~\fluxunit\ ,
implying an initial \tiff\ mass of $1.5\times10^{-4}$~\msun,
higher than for the assumption of a power law continuum.
Note that synchrotron models also predict a spectral steeping,
resulting in a similar \scff\ flux estimate.\cite{Vink01}

\section{New constraints on the average magnetic field in Cas A}
The relativistic electrons responsible for the radio synchrotron emission
from Cas A also give rise to gamma-ray bremsstrahlung and inverse Compton emission. 
The synchrotron emission for an electron energy 
power law distribution with index $q$, normalization  $K$, 
and magnetic field $B$ scales as 
$K B^{(q+1)/2} \nu^{-(q-1)/2}$.~\cite{Ginzburg65} 
Bremsstrahlung on the other hands scales with $K \Sigma_i n_i Z_i^2 $, 
with $n_i, Z_i$ the density and charge of ion $i$. 
Inverse Compton emission scales with 
the photon energy density and has a spectral slope that is
equal to that of synchrotron emission.
For Cas A the radio flux density for the epoch 2000.0
is 2522~Jy at 1 GHz with a spectral index of $\alpha = -0.78$, 
which is based on a compilation of radio measurements.

From the above it is clear that measuring bremsstrahlung or 
inverse Compton emission will constrain the average magnetic field 
strength.\footnote{More precisely, with bremsstrahlung we can determine 
$<KB^{(q+1)/2}>/<K \Sigma_i n_i Z_i^2 >$.} 
This method has been used to derive lower limits on
$B$ from upper limits on gamma-ray emission,
the latest estimate being $B > 0.35$~mG.\cite{Cowsik80,Atoyan00}
Here we apply this method to the hard X-ray emission,
using the \sax\ data in combination with archival 
\osse\ data (viewing periods 34 to 815). 
The \osse\ data are useful for setting  limits on the emission
above 100~keV.\cite{The96}

It turns out that limits on the inverse Compton emission does not
constrain the magnetic field significantly.
Taking into account the cosmic microwave 
background, synchrotron self Compton emission and 
Cas A's infrared emission, we estimate $B > 2\times 10^{-5}$~G.

Bremsstrahlung from the cosmic ray electrons is, however, much more constraining, 
but for the hard X-ray emission there is the caveat that the 
emitting electrons are only mildly relativistic and cannot be simultaneously 
observed at radio frequencies,
unlike electrons with $E \simgt 100$~MeV.
Even in the simplest model
we have to take into account that the shock acceleration
process produces a power law spectrum in momentum, producing a steepening in 
the energy spectrum from $E^{-\frac{1}{2}(q+1)}$ for $\gamma \simlt 2$ 
to $E^{-q}$ for $\gamma \gg 1$.\cite{Asvarov90}
More realistic models, incorporating the injection of electrons from the 
thermal plasma, predict a flattening of the energy spectrum from 
$E \sim kT_{\rm e}$ to $E \sim m_{\rm e}c^2$.~\cite{Bykov99}
In fact, it is possible that the hard X-ray emission is the result of this
injection spectrum.
Note, however, that we can still use the simpler model for obtaining
upper limits.
The reason is that a hard X-ray spectrum steeper than assumed here results in a lower 
electron cosmic ray normalization, $K$, for  $\gamma \gg 1$, 
resulting in a higher magnetic field.

The electron energy distribution scaling with $E^{-\frac{1}{2}(q+1)}$ produces
a spectrum with a spectral index of 2.2. The best fit normalization for such
a component is $(4.0\pm1.5)\times 10^{-7}$~ph\,s$^{-1}$keV$^{-1}$cm$^{-2}$ with
a 2$\sigma$ upper limit of 
$6.2\times 10^{-7}$~~ph\,s$^{-1}$keV$^{-1}$cm$^{-2}$.
Fitting a complete bremsstrahlung model gives 
$K \Sigma_i n_i Z_i^2 V/4\pi d^2 < 65$,
with $d$ the distance to Cas A and $V$ the emitting volume.
Soft X-ray observations imply $\Sigma_i n_i Z_i^2 \simeq 20$.\cite{Vink96}
Combining this with the expression for radio synchrotron 
emission\cite{Ginzburg65} gives $B > 2$~mG. 
This is comparable to, or even larger than estimates based on magnetic field
equipartition. Such a high magnetic field is necessary for the LHW model
to work, given the observed electron temperature.\cite{Laming01b}
Note that the upper limits on the bremsstrahlung component predict a 
continuum flux density an order of magnitude below 
the recent marginal continuum detection around 2~MeV by 
\comptel.\cite{Strong00}

\section{Future observations with \integral}
\integral, ESA's next major gamma-ray mission,  will cover the energy
range of 15 keV to 10 MeV and is planned for launch in 
October 2002.\cite{Winkler96}
Observations of nuclear line emission are one of its goals.
Cas A will be observed for 1.5 Ms by \integral\ in order to observe
\scff\ and \caff\ line emission with its two main instruments SPI and IBIS.
Other young supernova remnants that will be observed are
Tycho's supernova, SN 1987A and RX J0852.0-4622 (``Vela jr'').\cite{Iyudin98}

For Cas A the presence of \tiff\ is now well established,
but \integral\ will be able to obtain new science with SPI's ability
to detect Doppler shifts of the \caff\ line with an accuracy of
$\sim 300$~km/s, from which important explosion properties can be inferred.
\tiff\ is expected to be at the base
of the ejecta, implying a relatively low velocity.
However, in the case of SN1987A it was discovered that
\nifs, which also originates from deep inside the ejecta,
was partially mixed with the outer layers. 
Moreover, in Cas A iron rich knots have been found outside the ejecta 
shell, implying that some core material was ejected with
surprisingly high velocities.\cite{Hughes00}

IBIS is the most sensitive instrument for the \scff\ lines and the hard
X-ray continuum.
Although it will not dramatically improve
the continuum spectrum observed by \sax-PDS, the fact that the 
\scff\ line contribution will be better known allows for a better
measurement of continuum spectrum around 80~keV, where,
a spectral steepening is expected.
The coded mask design will ensure a better discrimination of possible
background sources.
\vskip 0.25cm

One of the other goals of \integral\ is the observation of
the electron-positron annihilation radiation, which has been observed
in the inner galaxy for a long time, most recently by \osse.\cite{Kinzer01}
The principle origin for the positrons is thought to be the decay of \scff\
and \cofs, the daughter of \nifs.
One might wonder whether it is possible to observe the annihilation radiation
directly in Cas A and other young supernova remnants.
Assuming for the annihilation time scale $\tau >> 300$~yr,
the expected 511~keV line  flux, $F_{511}$, can be expressed as:
\cite{Chan93}
\begin{equation}
F_{511} = (2 - 1.5f) \frac{N_{\beta^+}}{4\pi d^2 \tau} 
= (2- 1.5 f) \bigl( 0.19 e \frac{M_{56}}{\rm 56 amu} + 
0.94 \frac{M_{44}}{\rm 44 amu}\bigr) 
\frac{1}{4\pi d^2 \tau}, 
\end{equation}
where $f$ is the annihilation fraction due to the formation of positronium,
which is probably small for the hot plasma in Cas A,
$M_{44}$ and $M_{56}$ are the explosion yields of \tiff\ and $^{56}$Ni
respectively, $e$ is the escape fraction of $^{56}$Co generated positrons and
$d$ is the distance toward Cas A.
$\tau$\ depends on the electron density and is expected to be
$5\times10^{4}$ to $5\times10^{5}$ for electrons densities typical
for Cas, $10 - 100$~cm$^{-3}$.
We assume $M_{44} = 10^{-4}$\msun\ and $M_{56}= 0.07$ \msun.
It turns out that the detectability of the 511~keV emission from Cas A depends
critically on the escape fraction $e$.

\begin{figure}[t]
  \hbox{
	\parbox{95mm}{\psfig{figure=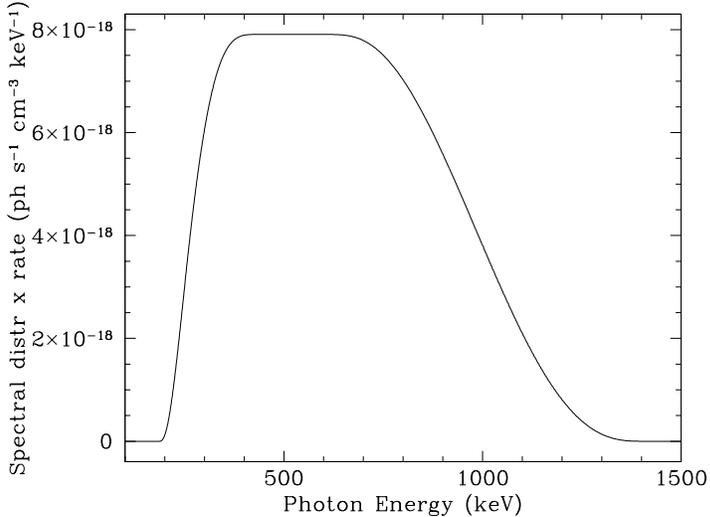,angle=-90,width=95mm}
	\vfill
	}
	\hskip 5mm
	\parbox{50mm}{
		\caption[]{
\label{fig:positrons}
Expected positron annihilation spectrum resulting 
from \scff\ and possibly \cofs\ decay positrons,
in the case of negligible positron energy losses.
The input positron spectrum was taken from ref. \cite{Chan93}.}}}
\end{figure}

Note that for $e = 0$, i.e. only \scff\ contributes, we obtain 
$F_{511} = 3\times10^{-6}$ at best. 
The average positron emitted by \scff\ has an energy of $\sim$600~keV, 
so the line may actually be significantly broadened to a continuum.
To our knowledge the emission spectrum of those hot positrons have never been 
published before, we therefore show it in Fig.~\ref{fig:positrons}.

If the 511~keV line is observed at all it must due to  a large escape
fraction of \cofs\ positrons. Calculations show typical escape fractions
of 1\% with possible fractions as high as 10\%.\cite{Chan93}
In the latter case, assuming $\tau = 4 \times 10^{4}$~yr, 
and if the positrons have slowed down considerably,
a flux as high as $3\times10^{-5}$\fluxunit\ might be expected, which is
observable by \integral. 
Although the prospects for such a high flux seems dim, it is certainly worth
looking for in Cas A and other supernova remnants. 

The observation of annihilation radiation will further improve
our understanding of the explosion. The positron escape
fraction depends sensitively on the explosion energy and asymmetries.
The reason is that early annihilation of positrons is slowed down if the
density in the expanding ejecta drops fast.
The annihilation in the ejecta depends on the positron energy losses,
mainly caused by ionizations.\cite{Chan93} For that reason also the composition
matters, e.g. ionization losses are smaller for He-rich ejecta.
Interestingly, most of the conditions that ensure a large escape fraction
are similar to the conditions for a high \tiff\ yield.
Note that high positron and gamma-ray escape fractions make a dim
supernova remnant. This is consistent with the lack of a bright supernova event
accompanying the explosion.

\section{Discussion}

We have discussed the recent detection of \scff\ line emission, which
supports the finding by \comptel\ of \caff\ emission and firmly establishes 
the presence of \tiff\ in Cas A.
Although there are some uncertainties in the flux values, caused by uncertainties
about the nature of the hard X-ray continuum, it nevertheless suggests that the
\tiff\ production was relatively high. This could be caused by a combination of
factors such as a highly asymmetric explosion or a very energetic explosion.
However, there are also some uncertainties in the modeling of supernovae preventing
a definitive conclusion on basis of \tiff\ alone.
The amount of \tiff\ synthesized indicates that at least 0.05~\msun of \nifs\ should 
have been synthesized, but this does not necessarily lead to a bright explosion as 
SN1987A, in which similar amounts of \tiff\ and \nifs\ were synthesized, 
was underluminous.
If Cas A was indeed an energetic explosion, and 0.07~\msun was synthesized, 
it may be
possible to detect 511~keV electron-positron annihilation emission. 
However, this is only possible under favorable circumstances, 
i.e. high annihilation rates now, 
and slow annihilation rates during the first few hundred days.

Future work, e.g. with \integral, will focus on the kinematics of 
\tiff\ and further
investigation of the nature of the hard X-ray continuum, 
which may be either synchrotron emission or bremsstrahlung.

\section*{Acknowledgments}
JV is supported by the NASA
%National Aeronautics and Space Administration 
through Chandra Postdoctoral Fellowship Award Number PF0-10011
issued by the Chandra X-ray Observatory Center, which is operated by the
Smithsonian Astrophysical Observatory for and on behalf of NASA under contract
NAS8-39073.
JML is supported by basic research funds of the Office of Naval Research.


\section*{References}
\begin{thebibliography}{99}

\bibitem{Ashworth80} W.B. Ashworth, \Journal{J. Hist. Ast.}{11}{1}{1980}
\bibitem{Fesen87} R. Fesen {\it et al.}, \Journal{\apj}{313}{37}{1987}
\bibitem{Vink96} J. Vink {\it et al.}, \Journal{\aap}{307}{L41}{1996}
\bibitem{Vink01} J. Vink {\it et al.}, \Journal{\apj}{560}{L79}{2001}
\bibitem{The98} L.-S. The {\it et al.}, \Journal{\apj}{504}{500}{1998}
\bibitem{Ahmad98} I. Ahmad {\it et al.}, \Journal{\PRL}{80}{2550}{1998}
\bibitem{Goerres98} J. G\"orres {\it et al.}, \Journal{\PRL}{80}{2554}{1998}
\bibitem{Norman98} E.B. Norman {\it et al.}, \Journal{\PRC}{57}{2010}{1998}
\bibitem{Diehl98} R. Diehl, \& F.X. Timmes, \Journal{\pasp}{110}{637}{1998}
\bibitem{Lundquist02} P. Lundqvist {\it et al.} \Journal{\aap}{374}{629}{2002}
\bibitem{Iyudin94} A. F. Iyudin {\it et al.}, \Journal{\aap}{284}{L1}{1994}
\bibitem{The96} L.-S. The {\it et al.}, \Journal{\aaps}{120}{357}{1996}
\bibitem{Rothschild97} R.E. Rothschild  {\it et al.}, \Journal{\NPB}{69}{68}{1997}
\bibitem{Vink00} J. Vink {\it et al.}, \Journal{25}{689}{\adspr}{2000}
\bibitem{Frontera97} F. Frontera {\it et al.}, \Journal{\aaps}{122}{357}{1997}
\bibitem{Schoenfelder00} V. Sch\"onfelder {\it et al.}, 
in ``Proc. of the 5th Compton Symposium'', McConnell \& Ryan, AIP Conf. Proc. 
{\bf 560}, 10 (2000)
\bibitem{Koyama95} K. Koyama {\it et al.}, \Journal{378}{255}{\nat}{1995}
\bibitem{Reynolds99} S.P. Reynolds \& J.W. Keohane, 
	\Journal{\apj}{525}{368}{199}
\bibitem{Asvarov90} A.I. Asvarov {\it et al.}, 
	\Journal{\aap}{229}{196}{1990}
\bibitem{Bykov99} A.M. Bykov \& Y.A. Uvarov, \Journal{\jetp}{88}{465}{1999}
\bibitem{Laming01a} J.M. Laming, \Journal{\apj}{546}{1149}{2001}
\bibitem{Laming01b} J.M. Laming, \Journal{\apj}{563}{828}{2001}
\bibitem{Bingham00} R.Bingham {\it et al.}, \Journal{\aaps}{127}{233}{2000}
\bibitem{Ginzburg65} V.L. Ginzburg \& S.I. Syrovatskii, \Journal{\araa}{3}{297}{1965}
%\bibitem{Rybicki} G.B. Rybicki \& A.P. Lightman, Radiative Processes in Astrophysics, Wiley, NY (1979)
\bibitem{Cowsik80} R. Cowsik \& S. Sarkar, \Journal{\mnras}{191}{855}{1980}
\bibitem{Atoyan00} A. Atoyan {\it et al.}, \Journal{\aap}{354}{915}{2000}
\bibitem{Strong00} A.W. Strong {\it et al.}, 
in ``Proc. of the 5th Compton Symposium'', eds. McConnell \& Ryan, AIP Conf. Proc. {\bf 560}, 10 (2000)
\bibitem{Winkler96} C. Winkler, \Journal{\aaps}{120}{637}{1996}
\bibitem{Iyudin98} A. Iyudin {\it et al.}, \Journal{\nat}{396}{142}{1998}
\bibitem{Hughes00} J.P. Hughes {\it et al.}, \Journal{\apj}{528}{L109}{2000}
\bibitem{Kinzer01} R.L. Kinzer {\it et al.}, \Journal{\apj}{559}{282}{2001}
\bibitem{Chan93} K.-L. Chan \& R.E. Lingenfelter, 
\Journal{\apj}{405}{614}{1993}
%\bibitem{Baars77} J.W.M. Baars {\it  et al.}, \Journal{\aap}{61}{99}{1977}

\end{thebibliography}
\end{document}